\documentclass[12pt]{article}

\usepackage[utf8]{inputenc}
\usepackage{graphicx}

\def\sign{\mathrm{sign}} 
\def\sh{\mathrm{sinh}}

\def\tr{\mathrm{tr}}

\title{Generating function for sine-Gordon correlators in finite volume from the inhomogeneous XXZ chain}
\author{F. Buccheri\footnote{mail address: buccheri@sissa.it}}
\date{}

\begin{document}

\maketitle
\maketitle
\begin{center}
 \textit{
SISSA and INFN, sezione di Trieste,
\\via Bonomea 265, 34136 Trieste, Italy}
\end{center}
\vspace{0.7cm}

\begin{abstract}
 We present an expression for the generating function of correlation functions in the sine-Gordon integrable field theory on a cylinder, with compact space. This is derived from the Destri-De Vega integrable lattice regularization of the theory, formulated as an inhomogeneous Heisenberg XXZ spin chain, and from more recent advances in the computations of spin form factors in the thermodynamic limit.
\end{abstract}

\section{Introduction}
Integrable massive field theories in two dimensions are studied since long time, serving as exactly solvable prototypes of strongly interacting theories.
Applications to problems in condensed matter and in statistical physics have been found over the years (see, e.g., \cite{EK04,Mussardo}) and range from off-critical statistical field theories to low-energy effective descriptions of conduction and magnetism in one dimension.

The sine-Gordon (SG) field theory is a paradigmatic example of integrable field theory because of both its simplicity and the richness of its features. In this paper, we are interested in computing SG correlation functions in Euclidean spacetime on a finite geometry, i.e. on a cylinder, on which the compactified direction will be regarded as space and the infinite one as time.

On the plane, the problem of computing correlation functions can be formally solved by applying the standard form factor approach \cite{KW78,Smirnov,L97}. Conversely,  for finite size, while studies of form factors have been carried on in semiclassical approximation in \cite{S98,MRS03}, the lack of a simple expression for the vacuum of the theory and its excitations still constitutes an obstacle for the method.

A way of tackling the problem may be that of interchanging the labels of space and time and consider the problem in which space is infinite, while the imaginary time is periodic, which amounts to considering the system at finite temperature. With this setup, it is possible to use the ``infrared`` form factors, but a suitable regularization is nonetheless necessary. One-point functions can be efficiently computed by series \cite{LM99,P11}; however, the formalism for two-point functions, perturbative in the exponential corrections in the volume, presently relies on a double series expansion containing regularized infinite-volume form factors \cite{PT10}, which results in cumbersome expressions for the model under consideration.

Going back to the original perspective, it is known \cite{L86a} that corrections to the particle masses which constitute the spectrum in the infrared limit are exponentially suppressed in the size of the system. This fact allowed the proposal \cite{PT08} of a finite-volume formalism, based on the infinite-volume form factors, that is correct up to terms which are exponentially suppressed in the volume. The method has been applied in \cite{FT11,FPT12} to soliton and breather form factors. An alternative regularization has been proposed in \cite{EK09} and applied to the computation of the dynamical structure factor in prototypical integrable field theories, although it is presently unclear how to extend the scheme to more general form factors.

The goal of this paper is that of deriving an exact expression for the generating function of correlation functions in finite volume, that includes exponential corrections in the size. The framework will be that of lattice integrable regularizations of the SG field theory and in particular of the one proposed in \cite{DDV87}. The expression derived will be written in terms of the counting function which solve the Destri-De Vega nonlinear integral equation (NLIE) as an expansion over the basis of the exact finite--volume eigenstates of the theory. We will moreover take advantage of the algebraic Bethe Ansatz results relative to the one-dimensional (inhomogeneous) Heisenberg magnet and in particular of the computation of the matrix elements of the magnetization operator in the limit \cite{KKMST09} in which the number of sites goes to infinity.

 By means of algebraic Bethe ansatz, a determinant representation for the generating function has been the goal of \cite{EFIK95}
. We believe that the subsequent advances on quantum spin chains (such as \cite{KMT98}) allow more explicit results. In connection with the inhomogeneous XXZ spin chain, the one-point functions of primary fields and their descendants in the sine-Gordon model have been analyzed in the framework of the recently explored fermionic structure of the model in \cite{JMS11b}, while the computation of form factors by separation-of-variables has been tackled in \cite{GMN12} recently.

Section \ref{sec:SGvsXXZ} reviews some basic facts about the field theory and its lattice regularization and allows us to fix the notation. The generating function of connected correlation functions is defined in \mbox{Section \ref{sec:generatingfn}} from the quantum field theory and from the spin chain perspectives and its expression is given in \mbox{Section \ref{sec:res}}. Manipulation of the lattice quantities is performed in the subsequent Section \ref{sec:latticecomputations}, then the scaling limit is performed in Section \ref{sec:scalinglimit}.

\section{Sine-Gordon and the inhomogeneous XXZ spin chain}\label{sec:SGvsXXZ}
The action of the sine-Gordon field theory on a cylinder of radius $L$ is:
\begin{equation}\label{SG_action}
\mathcal{S}=\int_{-\infty}^{\infty}d\tau \int_0^L dx \left[\frac{1}{2}\partial_\nu\phi\partial^\nu\phi-\frac{\mu^2}{\beta^2}\cos\left(\beta\phi\right) \right]
\end{equation}
where $\mu$ and $\beta$ are real parameters. In infinite volume, the fundamental excitations are known to be the soliton, with mass $m=m(\mu^2)$ and unit topological charge, and the antisoliton, with equal mass and opposite charge. A soliton and an antisoliton can bind together and form a breather, and breather scattering can lead to the production of a higher mass breathers as well. To be specific, bound states are labeled by integers and their number depends on the value of $\beta$. By defining the parameter $p=\frac{\beta^2}{8\pi-\beta^2}$, it is possible to distinguish two regimes: a \emph{repulsive} one, in which only the soliton and the antisoliton are present in the spectrum, and an \emph{attractive} one, in which a number $\lfloor 1/p\rfloor$ of bound states are allowed.

Correlation functions of the fields on the plane can be in principle generated by the knowledge of the form factors of the exponential of $\phi$, which have been computed in \cite{L93,L97}, while field form factors have been studied in \cite{Smirnov,BFKZ99,BK01}.

The lattice construction of \cite{DDV87}, based on a vertex model, can be formulated by means of an inhomogeneous XXZ spin chain, as underlined by the authors. Essentially, the fermionic variables defined on the bonds among vertexes, naturally identify spin variables via Jordan-Wigner transformation. The alternating sign rapidities which are associated to the bonds constitute the inhomogeneities of the XXZ chain, while the interaction among fermions of the original work is related to anisotropy parameter of the easy-axis interaction and to the coupling $\beta$ in the sine-Gordon Lagrangian. It has been shown in \cite{DDV97,FRT98a,FRT98b,FMQR97} that this formalism is able to reproduce many features of the theory in the limit $L\to \infty$ (energy spectrum, scattering phases) as well as those expected from conformal field theory (operator scaling dimensions) when $L\to 0$.

Consider \cite{DDV93} a lattice Heisenberg spin $1/2$ chain with $2N$ sites, characterized by local R-matrices of the form:
\begin{eqnarray}\label{Lax}
 &&R_{0,n}(\lambda)=\frac{1-c}{2}+\frac{1+c}{2}\sigma_0^z\sigma_n^z+\frac{b}{2}\left(\sigma_0^x\sigma_n^x+\sigma_0^y\sigma_n^y\right)  \nonumber\\
&& b=\frac{\sinh\lambda}{\sinh\left(i\gamma-\lambda\right)}\qquad c=\frac{\sinh i\gamma}{\sinh\left(i\gamma-\lambda\right)}
\end{eqnarray}
in which $\lambda$ is a complex ''Bethe ansatz'' rapidity and $\gamma$ is the anisotropy parameter, since in the homogeneous case is related to the third-component interaction of the spin degrees of freedom. It is related to the field theory parameters above via $p=\frac{\pi}{\gamma}-1$. The indexes $0$ and $n$ refer to the auxiliary $\mathbf{C}^2$ local Hilbert space and to the one relative to the $n-$th site.

The monodromy matrix for an inhomogeneous generalization of the XXZ chain can be built by means of an auxiliary $C^2$ space as follows:
\begin{eqnarray}\label{monodromy}
T(\lambda) =  e^{i\omega (\sigma_0^z-1)} R_{0,2N}(\lambda-\Lambda+i \gamma/2)R_{0,2N-1}(\lambda+\Lambda+i \gamma/2)\ldots R_{0,1}(\lambda+\Lambda+i \gamma/2)
\end{eqnarray}
in which twisted boundary condition with twist parameter $\omega$ have been introduced.

Eigenstates of the transfer can be written through the application of generalized creation operators $\mathcal{B}(\lambda)$ on a reference state, chosen here to be the one in which all the local spins are are polarized along the positive direction:
\begin{equation}\label{eigenstate}
\mathcal{B}(\mu_1)\ldots \mathcal{B}(\mu_M)\left|\uparrow\ldots\uparrow\right\rangle
\end{equation}
Each $\mathcal{B}$ operator lowers the total spin of the system by one, hence the total spin of the state above is given by 
\begin{equation}\label{spinchainspin}
S=N-M 
\end{equation}
In the following, the distinguished role of the field theory ground state will be played by the antiferromagnetic ground state, in which $N=M$ and the total spin is vanishing.

The set of Bethe rapidities $\{\mu\}$ satisfies the system of Bethe ansatz equations, which for this model can be written by defining a function of the variable $x$:
\begin{equation}\label{BaeX}
B(x|\{\mu\},\omega) = {B^{\omega}_\mu(x)} = \frac{a(x)}{d(x)}\prod_a\frac{\sinh(x-\mu_a+i\gamma)}{\sinh(x-\mu_a-i\gamma)}e^{2i\omega}
\end{equation}
 relative to the eigenstate identified by the roots $\{\mu\}$, with
\begin{eqnarray}\label{ad}
a(x)=\left[{\sinh(x-\Lambda-i\frac{\gamma}{2})\sinh(x+\Lambda-i\frac{\gamma}{2})}\right]^{M}  \nonumber\\
d(x)=\left[{\sinh(x-\Lambda+i\frac{\gamma}{2})\sinh(x+\Lambda+i\frac{\gamma}{2})}\right]^{M}
\end{eqnarray}
 Then the Bethe ansatz equations (BAE) are expressed like:
\begin{equation}\label{BAE}
B^{\omega}_\mu(\mu_j)=-1,\;\;j=1,\ldots,M
\end{equation}
It is possible from (\ref{monodromy}) to define a ``twisted'' transfer matrix, whose eigenvalue relative to the state $\left|\Psi_\omega(\{\lambda\})\right\rangle$ is:
\begin{equation}\label{transfermateigenvalues}
\tau_\omega(x|\{\lambda\})=
a(x)\prod_j\frac{\sinh(x-\lambda_j+i\gamma)}{\sinh(x-\lambda_j)}+
e^{-2i\omega}d(x)\prod_j\frac{\sinh(x-\lambda_j-i\gamma)}{\sinh(x-\lambda_j)}
\end{equation}

In the following, it will turn out to be useful the use of the rescaled variables (the ``rapidities'' of field theory) and inhomogeneities, and in particular of:
\begin{equation}\label{theta}
\theta_j=\frac{\pi}{\gamma}\mu_j
\end{equation}
The sine-Gordon model is then recovered by the double scaling
\begin{equation}\label{scalingDdV}
 \Theta=\frac{\pi}{\gamma}\Lambda\qquad\qquad \Theta\to\infty\quad,\quad a\to0\quad,\quad a\,e^{\Theta}=const=\frac{4}{m}
\end{equation}
with $m$ the soliton mass.

Because of the $i\pi$-periodicity of the Bethe equation, one is free to chose to restrict the rapidites to the fundamental strip as $\left]-\frac{\pi}{2},\frac{\pi}{2}\right]$, which corresponds, in terms of rescaled rapidities, to the strip $\left]-\frac{\pi}{2}(p+1),\frac{\pi}{2}(p+1)\right]$.

The logarithmic form of (\ref{BaeX}), i.e., of (\ref{BAE}) when considered as a function of one rapidity $\mu_j$, defines in this region the \emph{counting function}:
\begin{equation}\label{Z}
Z(x)=-i\log B(\frac{\gamma}{\pi}x)
\end{equation}
which takes its name from the fact that solutions of the logarithmic form of the Bethe equations are characterized by:
\begin{equation}\label{countingfnquantization}
 Z(\lambda_j)=2\pi I_{j}\qquad,\qquad I_{j}\in \mathbf{Z}+\frac{1+\delta}{2}\;,\qquad \delta=(N-S)_{mod2}
\end{equation}
for some integer or half-integer $I$, with $\delta=0,1$ specifying the sector of the field theory the state belongs to, as explained in, e.g., \cite{R00}.

Here we give only a short summary of the properties of the counting function, more can be found in \cite{DDV92,DDV93,DDV95,FMQR97,FRT98a,FRT98b,FRT98c,FRT99}. The function $Z$ is real analytic and satisfies the nonlinear integral equation (NLIE):
\begin{equation}\label{NLIE}
 Z(\theta)=m\,L\,\sh\theta+g(\theta|\left\{I\right\})-i\sum_{\sigma=\pm}\int dx G(\theta-x^\sigma)\log_{FD}\left(1+(-)^\delta e^{i\sigma Z(x^\sigma)}\right) +\alpha
\end{equation}
with $\alpha=\frac{p+1}{p}\omega$ for neutral states and $\omega<\pi/2$ and
\begin{eqnarray}
g(\theta|\left\{I\right\})&=&\sum_h^{N_H} \chi(\theta-\lambda_h)-\sum_c^{M_C} \chi(\theta-\lambda_c) \label{NLIEdefinitionssource}
\\
&&
-\sum_s^{N_S}\left(\chi(\theta-\hat y^-)+\chi(\theta-\hat y^+)\right)
-\sum_w^{M_W} \chi_{II}(\theta-\lambda_w)  \nonumber\\
\chi(x)&=&2\pi\int_0^\theta G(x)dx	 \label{NLIEdefinitionsChi}	  \\
G(\theta)&=& (p+1)\int\frac{dk}{2\pi}\frac{\sinh\frac{p-1}{2}k}{\cosh\frac{\pi k}{2}\sinh\frac{\pi p k}{2}}\cos(\theta k) \label{NLIEdefinitionsG}
\end{eqnarray}
where we denoted 
$x^\pm=x\pm i\eta$, with $\eta$ some small real quantity, and the subscript ``$II$'' reminds that the function $\chi$ must be considered in the second determination \cite{DDV93,DDV97,FRT98b} when the imaginary part of its argument exceeds $\min(1,p)\pi$, as in the case of wide roots. To clarify the definition of the source term $g$ above, we need to explain that solutions of (\ref{countingfnquantization}) can be classified according to their position in the complex plane as:
\begin{itemize}
 \item \emph{real solutions} of the equation
\begin{equation}\label{polesfactor}
1+(-1)^\delta e^{i\,Z(\lambda)}=0 
\end{equation}
constitute the Dirac sea in the field theory limit. They will be labeled by using a tilde on the variable and their set by $\{\tilde\lambda\}$.
 \item \emph{holes} are real solutions of (\ref{polesfactor}) that are not among the Bethe roots. Their number is $N_H$.
 \item \emph{special} solutions are real solutions of (\ref{polesfactor}) in which the counting function has a negative derivative. Their number is denoted by $M_S$.
 \item \emph{close} roots are present in pairs and have imaginary part in the strip between $-\min(p,1)\pi$ and $\min(p,1)\pi$. Their number is $M_C$.
 \item \emph{wide} roots are present in pairs and have the absolute value of the imaginary part in the strip between $\min(p,1)\pi$ and $\pi(p+1)/2$. There will be $M_W$ wide roots.
 \item \emph{self-conjugated} roots sit on the boundary of the periodicity strip and have imaginary part equal to $\frac{\pi}{2}(p+1)$.
\end{itemize}

The ground state of the sine-Gordon model is realized as the unique state which has all the roots on the real axis, quantized with all consecutive half-integer quantization numbers. In the scaling limit, excited states are completely specified by the quantum numbers of the holes and of the complex roots, whose combinations correspond to the IR excitations. The rapidities corresponding to these objects are determined according to (\ref{countingfnquantization}), and self-consistently determine the source term in (\ref{NLIE}), hence the counting function itself.

The total spin of an excited state above the antiferromagnetic ground state will be called, in field-theoretical language, topological charge $Q$ of the state itself. One can paraphrase the relation (\ref{spinchainspin}) for the total spin in terms of the excitations defined above as:
\begin{equation}\label{topologicalcharge}
Q=2S=N_H-M_C-2N_S-(1+\sign(p-1))M_W
\end{equation}
In general \cite{DDV97}, it can be stated that in the repulsive regime, each hole in the source terms carries a unit $U(1)$ charge, which in the language of spin chain corresponds to a unit spin, i.e., to a missing creation operator. Such a charge is lowered by one for every close root and by two for every wide root. On the other hand, in the attractive regime, wide roots correspond to independent excitations, not carrying any $U(1)$ charge. We then expect that the wide roots correspond to creation operators, therefore lowering the spin by one, only in the repulsive regime $p>1$. Conversely, in the attractive regime $p<1$, wide roots enter the expression of a state only through their effect on the other roots and the determination of a counting function.

With respect to the infrared description of the spectrum, it is known \cite{FRT98b} (see also \cite{Fth}) that 
\begin{itemize}
 \item \emph{soliton} and \emph{antisolitons} correspond to holes in the Fermi sea, quantized with half-integers.
 \item the solitons \emph{polarization} states are described by arrays of the first kind, having common real part and containing in any case exactly one pair of close roots. The position of the roots for asymptotically large volumes \cite{DDV93} can be determined with exponential precision in the size and is reported here below. Arrays can be either:
	\begin{itemize}
	\item odd degenerate
		\begin{equation}\label{array-I-odd}
		\theta_0=\theta+i\pi\frac{p+1}{2}\;,\qquad\theta_k=\theta\pm i\pi\left(\frac{1-p}{2}- k p\right)
					\qquad k=0,1,\ldots,\left\lfloor\frac{1}{2p}\right\rfloor
		\end{equation}
		with real $\theta$;
	\item even 
		\begin{equation}\label{array-I-even}
		\theta_k=\theta\pm i\pi\left(\frac{1}{2}- k p\right)	\qquad k=0,1,\ldots,\left\lfloor\frac{1}{2p}\right\rfloor
		\end{equation} 
 	\end{itemize}
\item the \emph{breather} degrees of freedom, when $p>1$, are described by arrays of the second kind, containing wide pairs only. These are, for $m\,L\to \infty$:
	\begin{itemize}
	\item odd degenerate
		\begin{equation}\label{array-II-odd}
		\theta_0=\theta+i\pi\frac{p+1}{2}\;,\qquad\theta_k=\theta\pm i\pi\left(\frac{1-p}{2}- k p\right)\qquad k=0,1,\ldots,s
		\end{equation} 
	\item even
		\begin{equation}\label{array-II-even}
		\theta_k=\theta\pm i\pi\left(\frac{1}{2}- k p\right)\qquad k=0,1,\ldots,s
		\end{equation} 
 	\end{itemize}
with $0\le s \le \frac{1}{2p}-1$ and real $\theta$. In particular, they describe the $2s+1$ and the $2s+2$ breather, respectively.
\end{itemize}

For simplicity, we will consider in the following only states in which the number of special objects is null and the counting function is monotonic on the real axis, which is the case for sufficiently large values of the size. With proper modification, the treatment can be in principle extended to account also for non monotonic counting functions, but this appears to be more cumbersome and will not be reported here.

The function $Z(\theta)$ is suitable for numeric computation and can be determined in a time of the order of minutes for the simplest root configurations. For a fairly larger amount of time, one can determine the counting function on a suitable grid on the complex plane, even if only its knowledge on three contours surrounding the real axis and the complex roots will be needed in the following.

\section{The generating function}\label{sec:generatingfn}
A convenient method for writing connected correlation functions is by differentiation of a generating function. In particular, we are interested in the expectation value on the exact finite-size vacuum $\Psi_0$:
\begin{equation}\label{expfield}
G_\omega(x)=\left \langle \Psi_0 \right| e^{-2i\pi\omega/\beta (\phi(x)-\phi(0))} \left| \Psi_0  \right\rangle_L
\end{equation}
in which $\omega$ is here a real number and the subscript $L$ stands for the size of the system. The field in the exponent is proportional to the fraction of topological charge in the interval $[0,x]$ (as in \cite{EFIK95}) and is realized on the lattice \cite{DDV87} by a string of operators acting on the local spin Hilbert spaces as:
\begin{equation}\label{string}
\left(\phi(x)-\phi(0)\right) = \frac{\beta}{2\pi} \sum_{l=1}^{2m}\sigma^z_l
\end{equation}
with $\sigma^z$ the usual Pauli matrix. A similar operator, the sum over projectors on the spin-up state, reads:
\begin{equation}\label{Q}
\mathcal{Q}_{2m+1}=\frac{1}{2}\sum_{l=1}^{2m}(1-\sigma^z_l)
\end{equation}

A convenient representation of the exponential of (\ref{Q}) was provided in \cite{KKMST09} in terms of the transfer matrices $\hat\tau_0, \hat\tau_\omega$ of two spin chains: one corresponding to the actual physical system, and the other to an analogous system in which a twist in the boundary condition $\omega$ had been introduced. 

We make use the results of \cite{KMT98}, which presents the solution of the inverse scattering for arbitrary inhomogeneities $\xi_l\;l=1,\ldots,2N$. The magnetization operator is written as:
\begin{equation}
\sigma^z_m=\prod_{l=0}^{m-1}\left(A+D\right)(\xi_l-i\gamma/2)\left(A-D\right)(\xi_m-i\gamma/2)\left(\prod_{l=0}^{m}\left(A+D\right)(\xi_l-i\gamma/2)\right)^{-1}
\end{equation}
which allows to write down \cite{KKMST09} the generating function in the inhomogeneous chain as:
\begin{eqnarray}\label{spinvert}
e^{-i\omega \sum_{l=0}^m \sigma_l^z} &=& 
\prod_{l=0}^{m} e^{-i\omega} \hat\tau_{\omega}\left((-1)^l\Lambda-i\gamma/2\right) \left(\prod_{l=1}^{m} {\hat\tau_{0}}\left((-1)^l\Lambda-i\gamma/2\right)\right)^{-1}
\end{eqnarray}
We will associate here the set of rapidities $\{\mu\}$ to the ground state and introduce a complete set of eigenstates of the twisted transfer matrix on the right of this operator. The transfer matrices act diagonally on the respective eigenstates, so that the expression obtained is:
\begin{equation}\label{latticeamplitudexpansion}
\frac{\left\langle \Psi_0 (\{\mu\}) \left| e^{i\omega\sum_j^m\sigma_j^z}\right|\Psi_0 (\{\mu\}) \right\rangle_L}
{\left\langle \Psi_0 (\{\mu\}) \Big|\Psi_0 (\{\mu\}) \right\rangle_L}
  = 
\sum_{\{\lambda\}_\omega}
\mathcal{A}(\{\lambda\})
\prod_{l=1}^m e^{-i\omega}\frac{\tau_{\omega}((-1)^l\Lambda-i\gamma/2|\{\lambda\}_\omega)}{\tau_{0}((-1)^l\Lambda-i\gamma/2|\{\mu\})}
\end{equation}
with
\begin{equation}\label{amplitudedef}
\mathcal{A}(\{\lambda\}_\omega) = \frac{\left|\left\langle \Psi (\{\lambda\}_\omega) |\Psi_0 (\{\mu\}) \right\rangle_L\right|^2}
{\left\langle \Psi_0 (\{\mu\}) |\Psi_0 (\{\mu\}) \right\rangle_L \left\langle \Psi (\{\lambda\}) |\Psi (\{\lambda\}) \right\rangle_L}
\end{equation}
The product on the right hand side yields, in the scaling limit, the phase:
\begin{equation}\label{translation2}
\prod_{l=1}^m e^{-i\omega}\frac{\tau_{\omega}((-1)^l\Lambda-i\gamma/2|\{\lambda\}_\omega)}{\tau_{0}((-1)^l\Lambda-i\gamma/2|\{\mu\})}
\to e^{-i x (\mathcal{P}(\{\lambda\}_\omega)-\mathcal{P}(\{\mu\}))}
\end{equation}
which is reviewed in \ref{translations}.

What written above is then a formal decomposition of the generating function of connected correlation functions. Its derivatives with respect to the twist $\omega$ provide a form factor expansion familiar in the framework of field theory. 
 For simplicity, we retain the form (\ref{latticeamplitudexpansion}), in which every term acts as a generating function for the vacuum-to-state probabilities, and refer to it as amplitude expansion.

The intermediate states are defined by the root structure which is encoded in the source term of (\ref{NLIE}), i.e., by the number of holes, close, ... roots and by their quantization numbers. The question about the completeness of the presently known solutions of the nonlinear integral equation is still open, to our knowledge.

The amplitudes have the property:
\begin{equation}\label{translation}
\mathcal{A}(\{\bar I\},x,t)=e^{-i\mathcal{P}(\{\lambda\})x+i(\mathcal{E}(\{\lambda\})-\mathcal{E}_0)t} \mathcal{A}(\{\bar I\},0,0)
\end{equation}
where $\mathcal{P}(\{\lambda\})$ is the total dressed momentum of the state $\Psi(\{\lambda\})$ and $\mathcal{E}(\{\lambda\})$ its energy, while $\mathcal{E}_0$ refers to the energy of the ground state.

We introduce here a compact notation which will be used throughout the paper:
\begin{equation}\label{shiftconvention}
 x^{\pm} = x \pm i \eta
\end{equation}
for some small $\eta$, denotes that the variable $x$ is slightly shifted above or below the real axis. 

Explicit expressions for the exact finite volume energy and momentum of a state identified by a counting function $Z$ and a given set of holes, complex and special roots are:
\begin{eqnarray}\label{DressedMomentum}
\mathcal{P}=\sum_{j}^{N_h} m\sinh h_j -\sum_{j}^{N_s}\left(m\sinh\,y_s^+ + m\sinh\,y_s^- \right)
 -\sum_{j}^{M_c} m\sinh c_j-\sum_{j}^{M_w} m\sinh w_j  \nonumber\\
-\frac{1}{\pi}\int_{-\infty}^{\infty} d\theta'\,\cosh\theta'\Im\log(1+e^{i Z(\theta'^+)})
\end{eqnarray}
\begin{eqnarray}\label{DressedEnergy}
 \mathcal{E}- \mathcal{E}_{bulk} = \sum_{j}^{N_h} m\cosh h_j -2\sum_{j}^{N_s} m\cosh\,y_s - \sum_{j}^{Mc} m\cosh c_j -\sum_{j}^{M_w} m\cosh w_j\nonumber\\
-\frac{1}{\pi}\int_{-\infty}^{\infty} d\theta'\,\sinh\theta'^+\Im\log(1+e^{i Z(\theta'^+)})
\end{eqnarray}
for which explicit computations can be found in \cite{Fth,FMQR97,DDV97}. The time shift phase comes from applying the double-row transfer matrix along the vertical direction along the lines of \cite{DDV93,F96lh}.

Note that the integrals in (\ref{DressedMomentum},\ref{DressedEnergy}) are well defined and finite with the convention (\ref{shiftconvention}). We underline here that, since the asymptotic behaviour of the counting function for large values of its argument is that of an hyperbolic sine, the factor 
$e^{\pm i Z(x^{\pm})}$ goes to zero faster than exponentially when $x\to\pm\infty$.

\section{The result}\label{sec:res}
Suppose the ket $|0\rangle_L$ to be the (finite volume) vacuum, which corresponds to the state in which all the roots lie on the real axis and are quantized by half-integers, without holes. The set of integers $\{\bar I\}$ defines instead an the excited state in the twisted system, as can be used as a starting point for solving self-consistently for the source terms in (\ref{NLIE}) and for the counting function itself. The sine-Gordon sector \cite{FRT98c} is reproduced by the configuration of roots having $2S+\delta+M_{sc}\in 2\mathbf{Z}$ and we will consider, for definiteness, half-integer quantization numbers for the rapidites, i.e. $\delta=0$. Hence, in the following, the number of self-conjugated roots is required to be even. To have a non vanishing matrix element, it is moreover necessary that the total number of roots in the excited state is the same as that of the ground state.

The generating function is given by
\begin{equation}\label{Gseries}
G_\omega(x) = N_\omega \sum_{\{I\}} e^{-i x (\mathcal{P}(\{I\})-\mathcal{P}_0) } \tilde{\mathcal{A}}(\{I\})
\end{equation}
where
\begin{equation}\label{Nseries}
  N_\omega = \left(\sum_{\{I\}} \tilde{\mathcal{A}}(\{I\})\right)^{-1}
\end{equation}

We shall now focus on the single terms of the series $\tilde{\mathcal{A}}$, which can be computed from the knowledge of the counting function of the ground state $Z_0$ and from the one for a generic twisted eigenstate $Z_\lambda$. Here below and in the following, we will denote the indexes relative to the holes of the excited states by $h$ and the ones relative to the complex roots, generically, by a $c$, so that a hole solution will be denoted by $\lambda_h$ and a complex root by $\lambda_c$. This shouldn't generate confusion with the ''close'' roots, as notation will be clear from the context. Moreover, to shorten notations, we write the signs $\{c_\alpha\}$, with the convention that $c_{holes}=1,\,c_{complex\,roots}=-1$.
We define hereby the functions:
\begin{equation}\label{varphi}
 \varphi_\rho(x,y)=\frac{mL\;\sinh\frac{\gamma}{\pi}(x-y)}{Z_{\rho}(x)-Z_\rho(y)}  \qquad \rho = \lambda,0
\end{equation}
\begin{equation}\label{Delta}
 \Delta(x)=\delta(x)-G(x)
\end{equation}
and:
\begin{eqnarray}\label{Deltalogratio}
\mathcal{L}_0^\sigma(x)&=&\int_{-\infty}^{\infty} du \Delta(x-u) \log(1+e^{i\sigma Z_0(u^{\sigma})})  \nonumber\\
\mathcal{L}_\lambda^\sigma(x)&=&\int_{-\infty}^{\infty} du \Delta(x-u) \log(1+e^{i\sigma Z_{\lambda}(u^{\sigma})})  \nonumber\\
\mathcal{L}^\sigma(x)&=&\int_{-\infty}^{\infty} du \Delta(x-u) \log\frac{1+e^{i\sigma Z_{\lambda}(u^{\sigma})}}{1+e^{i\sigma Z_0(u^{\sigma})}}
\end{eqnarray}

With these definitions, the expression for the term associated to the twisted state $\{\lambda\}$ and the vacuum, evaluated in the origin, is:
\begin{eqnarray}\label{amplitude}
\tilde \mathcal{A}(\{I\}) &=& 
\frac{\mathcal{S}\;\Phi \;\mathcal{D}}{\cosh^2\Sigma}
\prod_c \mathcal{C}(\lambda_c) \prod_h \mathcal{H}(\lambda_h) 
\mathcal{R}
\end{eqnarray}
in which products are taken over the complex roots $\lambda_c$ and the holes $\lambda_h$ that define the excited state and the quantities below appear.

The complex roots are taken into account by the factor:
\begin{eqnarray}\label{c-factor}
 \mathcal{C}(\lambda_c)=
\frac{\pi}{\gamma} \frac{\cos\frac{Z_0(\lambda_c)}{2}}{Z'_{\lambda}(\lambda_c)}
\exp\Big[-\frac{2\gamma}{\pi^2}  \int_{-\infty}^{\infty} du \Im\mathcal{L}_+(u) \coth\frac{\gamma}{\pi}(\lambda_c-u)\nonumber\\
 \qquad
+
 2\int_{-\infty}^{\infty} du \sum_\alpha c_\alpha G(\lambda_\alpha-u)\log\sinh\frac{\gamma}{\pi}(\lambda_c-u)
\Big]		
\end{eqnarray}
where the sum runs over the holes and complex roots $\lambda_\alpha$ with the convention stated above, with the usual understanding that the second determination has to be used for $G$ whenever $\lambda_\alpha$ is a wide root. The concerned reader may note that the logarithm of an hyperbolic sine grows linearly when its argument go to infinity, while the factor $G$ decreases exponentially, hence the last integral is convergent.
 The holes enter in the result by the factor:
\begin{eqnarray}\label{h-factor}
 \mathcal{H}(\lambda_h)= 
 \frac{\pi}{\gamma}\frac{\cos\frac{Z_0(\lambda_h)}{2}}{Z_{\lambda}'(\lambda_h)}
 \exp\Big[
-2\sign(1-p)M_w 
-2\int_{-\infty}^{\infty} du \sum_\alpha c_\alpha G(\lambda_\alpha-u)\log\varphi_\lambda(\lambda_h,u^-)
  \nonumber\\
\qquad
+2\int_{-\infty}^{\infty} \frac{du}{\pi} \Im \left(\mathcal{L}_0^+(u)\partial_x \log\varphi_0(\lambda_h,u^+)
 -\mathcal{L}_\lambda^+(u)\partial_x \log\varphi_\lambda(\lambda_h,u^+)\right)
\Big]
\end{eqnarray}
while the term $\mathcal{D}$ contains the various sources mixed:
\begin{eqnarray}
\mathcal{D}&=&\label{Dterm}
\frac{\prod_{c\ne c'}\sinh\frac{\gamma}{\pi}(\lambda_c-\lambda_{c'})
\prod_{h\ne h'}\sinh\frac{\gamma}{\pi}(\lambda_h-\lambda_{h'})}
{\prod_{ch}\sinh\frac{\gamma}{\pi}(\lambda_c-\lambda_h)\sinh\frac{\gamma}{\pi}(\lambda_c-\lambda_h)}
\end{eqnarray}
The factors that embody the contribution of the ``Fermi sea'' of Bethe roots in the thermodynamic limit are denoted by $\Phi$ and $\mathcal{S}$, the former being:
\begin{eqnarray}\label{Phifactor}
 \Phi &=&  \exp  \Big[ 
\int_{-\infty}^{\infty} dx \int_{-\infty}^{\infty} dy \sum_{\alpha,\beta} c_\alpha  c_\beta G(\lambda_\alpha-x) G(\lambda_\beta-y)\log\varphi_\lambda(x,y)-M_W^2 
\nonumber\\
&&
  -\sum_{\sigma,\sigma'=\pm}  \frac{\sigma\sigma'}{(2\pi)^2}
			\int_{-\infty}^{\infty} dx \mathcal{L}^{\sigma'}(x)
			\int_{-\infty}^{\infty} dy  \big[ \mathcal{L}_\lambda^\sigma(x)
			 \partial^2_{x,y}\log\varphi_\lambda(x^\sigma,y^{\sigma'}) - \mathcal{L}_0^\sigma(x)
			 \partial^2_{x,y}\log\varphi_0 (x^\sigma,y^{\sigma'})	\big]
 \nonumber\\
&&
 + \sum_{\alpha} c_\alpha \int_{-\infty}^{\infty} \frac{dx}{\pi} G(\lambda_\alpha-x) \int_{-\infty}^{\infty} dy 
 \Im \left[ \mathcal{L}_\lambda^{+}(x)\partial_y \log\varphi_\lambda(x,y^{+})
			-\mathcal{L}_0^{+}(x)\partial_y \log\varphi_0(x,y^{+})\right]
  \Big]
\nonumber\\
\end{eqnarray}
again with the same conventions on the sums over sources. Note once again that the function $\log\varphi_\lambda$ is linear in $x$ or $y$ when these are large, therefore the kernel $G$ ensures convergence of the integrals in the first line. We also have the factor:
\begin{eqnarray}\label{Csfactor}
\mathcal{S} &=&\exp\Big[-
\frac{\gamma^2}{\pi^2}\sum_{\sigma\sigma'=\pm}\frac{\sigma\sigma'}{4\pi^2}\int_{-\infty}^{\infty} dx 
\int_{-\infty}^{\infty} dy \mathcal{L}_\sigma(x)\mathcal{L}_{\sigma'}(y)
\frac{1}{\sinh^2\frac{\gamma}{\pi}(x^{\sigma}-y^{\sigma'}-i\pi)}
 \nonumber\\
&&\qquad
 + \int_{-\infty}^{\infty} dx \int_{-\infty}^{\infty} dy \hat \Delta(x) \hat \Delta(y) \log\sinh\frac{\gamma}{\pi}(y-x-i\pi)
 \nonumber\\
&& 
 + \frac{\gamma}{\pi}\sum_{\sigma=\pm}\frac{\sigma}{2\pi i}\int_{-\infty}^{\infty} dx \int_{-\infty}^{\infty} dy
	\hat\Delta(x)\mathcal{L}_\sigma(y)
     \frac{\sinh\frac{\gamma}{\pi}(y^\sigma-x)}{\sinh\frac{\gamma}{\pi}(y^\sigma-x-i\pi)\sinh\frac{\gamma}{\pi}(y^\sigma-x+i\pi)}
\Big] \nonumber\\
\end{eqnarray}
where we have made use of the notation $\hat\Delta$ to denote:
\begin{equation}\label{deltahat}
\int_{-\infty}^{\infty} du\hat \Delta(u) f(u)=\sum'_{\alpha}c_\alpha f(\lambda_\alpha) - \sum_{\alpha}c_\alpha\int_{-\infty}^{\infty} du G(\lambda_\alpha-u)f(u)
\end{equation}
and the sum is over all the sources, but the prime excludes the wide roots from the sum if $p<1$. We would like to recall once again the fact that the integrations over variables with a superscript are not on the real axis, according to or convention (\ref{shiftconvention}).
 The argument of the hyperbolic cosine in (\ref{amplitude}) is given by
\begin{equation}\label{coshfactor}
 \Sigma = -\sum_\alpha \left(\lambda_\alpha- \int_{-\infty}^{\infty} du \, u\, G(\lambda_\alpha-u) \right) 
+ \frac{p+1}{2p} \Im \int_{-\infty}^{\infty} \frac{du}{\pi}
\log\frac{1+e^{i Z_\lambda(u^+)}}{1+e^{i Z_0(u^+)}}
\end{equation}
and once again the first integration is convergent due to the asymptotic behaviour of $G$ and the second due to the one of the counting function.
%
The last term $\mathcal{R}$ is a ratio of determinants:
\begin{equation}
\mathcal{R}=\label{Rterm}
\left|
\frac{\det\left[1-\hat W_{0,\lambda}\right]\det\left[1+\left( A_+-1\right)\hat G_{-\omega}\right]}{\det\left[1-\hat W_{0}\right]}
\frac{\det\left[1-\hat W_{\lambda,0}\right]\det\left[1+\left(A_+^{-1}-1\right)\hat G_{\omega}\right]}{\det\left[1-\hat W_{\lambda}\right]}
\right|
 \end{equation}

%
%
Now we introduce the integral operators appearing in the overlap determinant. They depend either on two complex variables $w,v$ and on two species indexes $\sigma,\sigma'=\pm$ or on the rapidities that define the excitations. They read:
\begin{eqnarray}\label{kernellambdamu}
W_{\lambda,0}^{\sigma,\sigma'}(w,v)&=&
\frac{1}{2\pi}\frac{A(w^\sigma)}{1+e^{-i\sigma Z_0(w^{\sigma})}}
\left(
G_{-\omega}\left(w^\sigma-v^{\sigma'}\right)
+ F_{-\omega}^{\sigma,\sigma'}\left(w,v\right)
\right) 
  \nonumber\\
W_{\lambda,0}^{\sigma'}(\lambda_c,v)&=&
 \frac{Res A(\lambda_c)}{1+e^{iZ_0(\lambda_c)}}
\left(
G_{-\omega}\left(\lambda_c-v^{\sigma'}\right)  
+ F_{-\omega}^{\sigma'}\left(\lambda_c,v\right)
\right) 
\nonumber\\
W_{\lambda,0}^{\sigma}(w,\lambda_c)&=&
\frac{1}{2\pi}\frac{A(w^{\sigma})}{1+e^{-i\sigma Z_0(w^\sigma)}}
\left(
G_{-\omega}\left(w^{\sigma}-\lambda_c\right)
 + F_{-\omega}^{\sigma}\left(w,\lambda_c\right)
\right)
  \nonumber\\
W_{\lambda,0}(\lambda_c,\lambda_{c'})&=&
 \frac{Res A(\lambda_c)}{1+e^{iZ_0(\lambda_c)}}
\left(
G_{-\omega}\left(\lambda_c-\lambda_{c'}\right)
+ F_{-\omega}\left(\lambda_c,\lambda_{c'}\right)
\right)
\end{eqnarray}
\begin{eqnarray}\label{kernelmulambda}
W_{0,\lambda}^{\sigma,\sigma'}(w,v)&=&
\frac{1}{2\pi}\frac{A^{-1}(w^{\sigma})}{1+e^{-i\sigma Z_{\lambda}(w^{\sigma})}}
\left(
G_{\omega}\left(w^{\sigma}-v^{\sigma'}\right)
 + F_{\omega}^{\sigma,\sigma'}\left(w,v\right)
\right)
 \nonumber\\
W_{0,\lambda}^{\sigma'}(\lambda_h,v)&=&
-\frac{A^{-1}(\lambda_h)}{Z'_{\lambda}(\lambda_h)}
\left(
G_{\omega}\left(\lambda_h-v^{\sigma'}\right)
+ F_{\omega}^{\sigma'}\left(\lambda_h,v\right)
\right)
  \nonumber\\
W_{0,\lambda}^{\sigma}(w,\lambda_h)&=&
\frac{1}{2\pi}\frac{A^{-1}(w^\sigma)}{1+e^{-i\sigma Z_\lambda(w^{\sigma})}}
\left(
G_{\omega}\left( w^{\sigma}-\lambda_h \right)
+ F_{\omega}^{\sigma}\left(w,\lambda_h\right)
\right)
  \nonumber\\
W_{0,\lambda}(\lambda_h,\lambda_{h'})&=&-
\frac{A^{-1}(\lambda_h)}{Z'_{\lambda}(\lambda_h)}\left(
G_{\omega}\left(\lambda_h-\lambda_{h'}\right)
+ F_{\omega}\left(\lambda_h,\lambda_{h'}\right)
\right)
\end{eqnarray}
and
\begin{equation}\label{Aplus}
A(w+i\pi/2)=A_+(w)=
e^{-i(Z_\lambda(w)-Z_0(w))-2i\omega}
\end{equation}
for which another expression is provided in Section \ref{sec:dets}. 

The determinants are of the Fredholm type, integrals are performed on the real axis and the the species indexes and the excitations variables are summed over as well. In facts, complex roots of the bra state must be explicitly summed over as well as holes of the ket, if any, subtracted, as exemplified later in (\ref{Wsquare}).
 The source function needs, for some configurations, to be evaluated in regions in which the imaginary part of the argument exceeds $\min(p,1)\pi$: it is therefore necessary to use the second determination \cite{DDV93,DDV97,FRT98b}. 

The integral operator $G_\omega$ is defined as:
\begin{equation}\label{Gomega}
G_\omega(w)=\intop_{-\infty}^\infty \frac{dk}{2\pi} e^{i\,\gamma k\,w/\pi}
\frac{\sinh\left[\left(\frac{\pi}{2}-\gamma\right)k+i\omega \right]}
{e^{-i\omega}\sinh\frac{\pi k}{2}+\sinh\left[\left(\frac{\pi}{2}-\gamma\right)k+i\omega \right]}
\end{equation}
and reduces to (\ref{NLIEdefinitionsG}) for $\omega \to 0$. Moreover,
\begin{eqnarray}\label{Fomega}
F_\omega^{\sigma,\sigma'}(w,v) &=& \sum_{n=1}^\infty \int_{-\infty}^{\infty} dx_1\ldots dx_n G_\omega\left(w^{\sigma}-x_1^{+}\right) \left(1-A(x_1^{+})\right) G_\omega\left(x_1-x_2\right)\ldots \nonumber\\
  && \qquad\ldots\left(1-A(x_n^{+})\right) G_\omega\left(x_n^{+}-v^{\sigma'}\right)
\end{eqnarray}
with and obvious extension to the case where one or both rapidities in the argument appear in the source.

For what the norm determinants are concerned, their expression can be written as:
\begin{equation}\label{normdet}
\det\left[1-\hat W_x\right] 
\end{equation}
where $x$ stands for one of the two states and:
\begin{eqnarray}
W_{0}^{\sigma,\sigma'}(w,v)&=&
\frac{1}{2\pi}\frac{1}{1+e^{-i\sigma Z_0(w^{\sigma})}}
G\left( w^{\sigma}-v^{\sigma'} \right)
 \nonumber\\
W_{\lambda}^{\sigma,\sigma'}(w,v)&=&
\frac{1}{2\pi}\frac{1}{1+e^{-i\sigma Z_\lambda(w^{\sigma})}}
G\left(w^{\sigma}-v^{\sigma'}\right)
 \nonumber\\
W_{\lambda}^{\sigma'}(\lambda_h,v)&=&
-\frac{1}{Z_{\lambda}'(\lambda_h)}
G\left(\lambda_h-v^{\sigma'}\right)
  \nonumber\\
W_{\lambda}^{\sigma}(w,\lambda_h)&=&
\frac{1}{2\pi}\frac{1}{1+e^{-i\sigma Z_\lambda(w^{\sigma})}}
G\left(w^{\sigma}-\lambda_h\right)
  \nonumber\\
W_{\lambda}(\lambda_h,\lambda_{h'})&=&-
\frac{1}{Z_{\lambda}'(\lambda_h)}
G\left(\lambda_h-\lambda_{h'}\right)
\end{eqnarray}
where the ``holes`` terms are present for a generic excited state and $\pi^-$ is a number slightly smaller than $\pi$.

 An interpretation in terms of pseudoparticles is possible: the finite-size vacuum can be written in terms of the fundamental excitations (solitons and antisolitons) of the infrared theory, which occupy the available levels according to a (complex) filling fraction containing the vacuum and excited pseudoenergies. Excitations constructed upon such a vacuum interact both among them and with the background pseudoparticles and the matrix elements of the operators show such features of the theory.

\section{Scalar products and norms}\label{sec:latticecomputations}
We need to perform the computation of the scalar products of the ground state with a generic ``twisted'' state \cite{S89,KMT98}. Let $\mu_1,\dots,\mu_M$ satisfy the system (\ref{BAE}), with twist $\omega$ to retain full generality and $\lambda_1,\dots,\lambda_M$ be generic complex numbers. Then
\begin{equation}\label{scal-prod}
\langle 0|\prod_{j=1}^{M}C(\lambda_j)|\psi(\{\mu\})\rangle =
 \frac{\prod_{a=1}^{M} d(\mu_a)}
{\prod\limits_{a>b}^M\sinh(\mu_a-\mu_b)\sinh(\lambda_b-\lambda_a)}
\cdot \det H(\{\mu\},\{\lambda\})
\end{equation}
in which $H$ is 
\begin{equation}
  H_{j k}^{\omega}= a(\lambda_j)\,t(\mu_k,\lambda_j)\,\prod_{l=1}^{M} \sinh(\mu_l-\lambda_j-i\gamma)
- e^{-2i\omega} d(\lambda_j)\,t(\lambda_j,\mu_k)\,\prod_{l=1}^{M} \sinh(\mu_l-\lambda_j+i\gamma)			
\end{equation}
with  $1\le j,k \le M$ and 
\begin{equation}\label{def-t}
t(\mu,\lambda)=\frac{-i\sin\gamma}{\sinh(\mu-\lambda)\sinh(\mu-\lambda-i\gamma)}.
\end{equation}

From this expression, we can extract both the overlaps and the norms of the states after some manipulation, in which we make explicit use that the rapidities $\{\lambda\}$ also satisfy (\ref{BAE}).

Two alternative expressions, which are suitable for the scaling limit, can be provided for the overlap. We refer the reader to \cite{KKMST09}, where the determinant of the overlap matrix was written as a Fredholm determinant on a contour.
With some variation of their method, in which use of the Bethe equations and of the definition (\ref{BaeX}) is explicitly made and whose details can be found in appendix  \ref{app:contoursum}, the scalar product (\ref{scal-prod}) can be written in a form which is more suitable for subsequent manipulation. In the following expressions, the quantity $\omega$ denotes the relative twist of the state.
\begin{eqnarray}\label{scal-prod2}
&&
\langle \psi(\{\lambda_j\})|\psi(\{\mu\})\rangle = \nonumber\\
&&
\qquad\qquad\quad
=  \frac{ e^{-2i\omega M}\prod_j d(\lambda_j) d(\mu_j) \left(1+B_\mu(\lambda_j)\right)}
{\cosh(\sum\lambda_l-\sum\mu_l)}
\prod_{j,k}\frac{\sinh(\mu_j-\lambda_k+i\gamma)}{\sinh(\mu_j-\lambda_k)}
       \det\left(1-\hat U^{-\omega} \right)
\nonumber \\
&& 
\qquad\qquad\quad
 = \frac{ \prod_j d(\lambda_j) d(\mu_j) \left(1+B_\lambda(\mu_j)\right)}
{\cosh(\sum\lambda_l-\sum\mu_l)}
\prod_{j,k}\frac{\sinh(\lambda_j-\mu_k+i\gamma)}{\sinh(\mu_j-\lambda_k)}
       \det\left(1-\hat U^{\omega} \right)
\end{eqnarray}
with the matrix
\begin{eqnarray}\label{Udef}
 U_{j,k}^{-\omega}&=&\frac{K_{-\omega}(\lambda_j-\lambda_k)}{1+B_\mu(\lambda_j)}
 \frac{\prod_l\sh(\lambda_j-\mu_l)}{\prod_{l\ne j}\sh(\lambda_j-\lambda_l)}
 \prod_{l}\frac{\sh(\lambda_j-\lambda_l-i\gamma)}{\sh(\lambda_j-\mu_l-i\gamma)}
\\
 U_{j,k}^{\omega}&=&\frac{K_\omega(\mu_j-\mu_k)}{1+B_\lambda(\mu_j)}
 \frac{\prod_l\sh(\mu_j-\lambda_l)}{\prod_{l\ne j}\sh(\mu_j-\mu_l)}
 \prod_{l}\frac{\sh(\mu_j-\mu_l-i\gamma)}{\sh(\mu_j-\lambda_l-i\gamma)}
\end{eqnarray}
and the function:
\begin{equation}\label{Komegadef}
 K_\omega(x) = \coth\frac{\gamma}{\pi}(x-i\pi)-e^{2i\omega}\coth\frac{\gamma}{\pi}(x+i\pi)\;,\qquad K_0(x)=K(x)
\end{equation}

For the computation of norms, one considers the limit $\{\lambda\},\{\mu\}\to\{\nu\}$, for which the matrix above becomes simply:
\begin{eqnarray}\label{simpleW}
U_{j,k}^{\omega_\nu}&=&\frac{K(\nu_j-\nu_k)}{1+B^{\omega_\nu}_{\nu}(\nu_j)}
\end{eqnarray}

For the remaining part of the section, we shall be using rescaled rapidity variables. Moreover, unless otherwise specified, we shall consider the state $\mu$ to be the ground state of the (untwisted) inhomogeneous chain, while the state $\{\lambda\}$ is considered as having a twist. 

 As a preliminary step, one observes that by applying the definition of counting function and a representation of the cosine as an infinite product:
\begin{equation}
1+B_{\mu}^\omega(x) =
2e^{-i/2\;Z_\mu(x)}\prod_{k=-\infty}^{\infty}\left(1-\frac{Z_\mu(x)}{2\pi(k-1/2)}\right)
\end{equation}
from which, considering the ground state with $2M$ roots $\{\mu\}$, having labels ranging on half-integers between $-M+1/2$ and $M-1/2$, we have:
\begin{equation}	\label{cosinetrick}
1+B_{\mu}^\omega(x)= 2e^{-i/2\;Z_\mu(x)}\prod_{I=-M+1/2}^{M-1/2}\left(\frac{Z_\mu(\mu_I)-Z_\mu(x)}{ Z_\mu(\mu_I)}\right)
\frac{\Gamma(M+\frac{1}{2})^2}{\Gamma(M+\frac{1}{2}-\frac{Z_\mu}{2\pi})\Gamma(M+\frac{1}{2}+\frac{Z_\mu}{2\pi})}
\end{equation}
The last ratio tends to unity in the limit in which the number of roots goes to infinity and will not be rewritten in the following. The case in which there is a finite number of excitations yields the same result, if one considers in the product above the set $\{\tilde \mu\}$ of all the real roots of (\ref{polesfactor}).

Having established this fact, we consider the state $\{\lambda\}$ to be excited and the state $\{\mu\}$ to be the vacuum, identified by the subscript $0$. We are moreover interested in the normalized matrix elements, so we divide the overlap by the norm of the two states.

Let us multiply and divide by the holes and the complex roots, in order to obtain expressions in which all and only the \emph{real solutions} appear. This is convenient in that we can consider the ratio between each hyperbolic sine appearing in the denominator of the expressions (\ref{scal-prod2}) and the differences of the counting function computed at the points in the argument of the sine, as arising from the product representation (\ref{cosinetrick}). Following \cite{KKMST09} this defines the functions (\ref{varphi}).

After illustrating the general procedure, it is simpler to consider two additional states, whose rapidities we label by $\{\rho\},\,\{\nu\}$: at the end of the computations, we will send $\{\rho\}\to\{\lambda\}$ and $\{\nu\}\to\{\mu\}$ and show that the poles arising from the factors of the kind (\ref{polesfactor}) are canceled by the zeros of the hyperbolic sines in the expression for the scalar product. In order to obtain a product involving only the real solutions, we consider the ratio:
\begin{eqnarray}\label{mansinhprod}
&&
\prod_{j,k}\frac{\sinh\frac{\gamma}{\pi}(\lambda_j-\rho_k)\sinh\frac{\gamma}{\pi}(\mu_j-\nu_k)}
{\sinh\frac{\gamma}{\pi}(\lambda_j-\nu_k)\sinh\frac{\gamma}{\pi}(\mu_j-\rho_k)} =
\frac{\prod_{cc'}\sinh\frac{\gamma}{\pi}(\lambda_c-\rho_{c'})\prod_{h,h'}\sinh\frac{\gamma}{\pi}(\lambda_h-\rho_{h'})}
{\prod_{ch}\sinh\frac{\gamma}{\pi}(\lambda_c-\rho_h)\sinh\frac{\gamma}{\pi}(\rho_c-\lambda_h)}
		\nonumber\\
&& \qquad
\prod_{j,k}\frac{\sinh\frac{\gamma}{\pi}(\tilde\lambda_j-\tilde\lambda_k)\sinh\frac{\gamma}{\pi}(\mu_j-\nu_k)}
{\sinh\frac{\gamma}{\pi}(\tilde\lambda_j-\mu_k)\sinh\frac{\gamma}{\pi}(\tilde\rho_j-\mu_k)}
\prod_h\prod_j\frac{\sinh\frac{\gamma}{\pi}(\lambda_h-\mu_j)\sinh\frac{\gamma}{\pi}(\rho_h-\mu_j)}
{\sinh\frac{\gamma}{\pi}(\lambda_k-\tilde\rho_j)\sinh\frac{\gamma}{\pi}(\rho_h-\tilde\lambda_j)}
		\nonumber\\
&& \qquad
\prod_c\prod_j\frac{\sinh\frac{\gamma}{\pi}(\lambda_c-\tilde\rho_j)\sinh\frac{\gamma}{\pi}(\rho_c-\tilde\lambda_j)}
{\sinh\frac{\gamma}{\pi}(\lambda_c-\mu_j)\sinh\frac{\gamma}{\pi}(\rho_c-\mu_j)}
\end{eqnarray}
On the other hand, for what the factor involving the counting function is concerned, we can write:
\begin{eqnarray}\label{decompref}
\prod_j\frac{1+e^{iZ_0(\rho_j)}}{1+e^{iZ_\lambda(\rho_j)}}=
\prod_c\frac{1+e^{iZ_0(\rho_c)}}{1+e^{iZ_\lambda(\rho_c)}}\prod_h\frac{1+e^{iZ_0(\rho_h)}}{1+e^{iZ_\lambda(\rho_h)}}
\prod_h\left(\frac{1+e^{iZ_\lambda(\rho_h)}}{1+e^{iZ_0(\rho_h)}}\right)^2
\prod_j\frac{1+e^{iZ_0(\tilde\rho_j)}}{1+e^{iZ_\lambda(\tilde\rho_j)}}
\end{eqnarray}
By multiplying the first and the second product with the first ratio of (\ref{mansinhprod}) and taking the limit to coinciding states, we obtain the term (\ref{Dterm}), which already contains a finite number of rapidities, apart from a phase factor.

The factors in the first term that contain the same index for the hole, together with the third term of (\ref{mansinhprod}) and the third term of (\ref{decompref}) yield:
\begin{eqnarray}\label{holefacprod}
\prod_h\mathcal{H}(\lambda_h)\quad,\qquad \mathcal{H}(\lambda(h))&=&\frac{1+e^{iZ_\mu(\lambda_h)}}{iZ_\lambda'(\lambda_h)}
\prod_j\frac{\varphi_0(\lambda_h-\tilde\mu_j)^2}{\varphi_\lambda(\lambda_h-\tilde\lambda_j)^2}
\end{eqnarray}
while the last of (\ref{decompref}), accompanied by the corresponding product in the $\mu$ rapidities, with the second in (\ref{mansinhprod}) provide the factor:
\begin{equation}\label{Phifactordef}
\Phi = \frac{\varphi_0(\mu_j,\mu_k)\varphi_\lambda(\tilde\lambda_j,\tilde\lambda_k)}{\varphi_0(\mu_j,\tilde\lambda_k)\varphi_\lambda(\mu_j,\tilde\lambda_k)}
\end{equation}
From all the previous expressions, one also obtains a phase factor containing sum over rapidities of the difference of the two counting functions, that will be of no relevance in the following.

The last product in (\ref{mansinhprod}) is already in a form suitable for the scaling limit; together with the part of the product in the first term that contains the same index for the close root, it may be rewritten as:
\begin{equation}\label{complfacprod}
\prod_c\mathcal{C}(\lambda_c) \quad,\qquad
\mathcal{C}(\lambda_C)=\frac{1+e^{iZ_\mu(\lambda_c)}}{iZ_\lambda'(\lambda_c)}
			 \left(\prod_j\frac{\sinh\frac{\gamma}{\pi}(\lambda_c-\tilde\lambda_j)}
{\sinh\frac{\gamma}{\pi}(\lambda_c-\tilde\mu_j)}\right)^2
\end{equation}
and constitutes a multiplicative contribution from complex roots.

 According to our previous analysis, we write:
\begin{equation}\label{normalized_matrix_element}
  \frac{\left|\langle\Psi(\{\lambda\})|\Psi(\{\mu\})\rangle\right|^2}
{{\left\|\Psi(\{\mu\})\right\|^2\left\|\Psi(\{\lambda\})\right\|^2}}
=
\mathcal{S} \; \mathcal{D} \; \Phi \; \prod_h \mathcal{H}(\lambda_h) \prod_c\mathcal{C}(\lambda_c)
 \left|\frac{\det\left(1-U^{\omega}_{\lambda,\mu}\right)\det\left(1-U^{-\omega}_{\mu,\lambda}\right)}
{\det\left(1-U_\lambda\right)\det\left(1-U_\mu \right)}\right|
\end{equation}
Where the definitions (\ref{Dterm}) and
\begin{eqnarray}\label{Sfactordef}
\mathcal{S} &=&\prod_{j,k}\frac{\sinh(\mu_j-\lambda_k-i\gamma)\sinh(\lambda_j-\mu_k-i\gamma)}
{\sinh(\mu_j-\mu_k-i\gamma)\sinh(\lambda_j-\lambda_k-i\gamma)}
\end{eqnarray}
have been used.

\section{The scaling limit}\label{sec:scalinglimit}
Assuming that a function $f$ is analytic in a connected region of the complex plane, that contains all the roots $\{\lambda\}$, satisfying (\ref{BAE}), then the sum of the values of the function when evaluated in this points can be written as
\begin{eqnarray}\label{sums}
\sum_j f(\lambda_j)& =& \int_{-\infty}^{\infty} \frac{du}{2\pi}\left(\frac{N}{\cosh(u-\Theta)}+\frac{N}{\cosh(u+\Theta)}\right)f(u) \nonumber\\
&&-\sum_{k=1}^{N_h} (\Delta\star f) (h_k)+\sum_{k=1}^{M_c} (\Delta\star f) (c_k)+\sum_{k=1}^{M_w}(\Delta_{II}\star f) (w_k)
 +2\sum_{k=1}^{M_s}(\Delta\star f)(s_k)\nonumber\\
&&+i\sum_{\sigma=\pm} \sigma \int_{-\infty}^{\infty} \frac{du}{2\pi}(\Delta\star f)(u^{\sigma})\log_{FD}'\left(1+e^{i\sigma Z(u^{\sigma})}\right)  
\end{eqnarray}
where the symbol $\star$ represents convolution over the real axis and the distribution (\ref{Delta}) appears. By using this expression, which can be obtained in an analogue way to the NLIE itself \cite{DDV95,DDV97,FMQR97,FRT98b}, one can compute the scaling limit of the factors above.

\subsection{The determinants} \label{sec:dets}
We want to reduce the expressions above to a standard Fredholm determinant form $\det(1+K)=e^{\sum_n (-1)^{n-1}\tr[K^n]/n}$. For an analytic function (at least on the real axis), one can write
\begin{equation}\label{smallsum}
\sum_j \frac{f(\lambda_j)}{\prod_{l\ne j}(\lambda_j-\lambda_l)}=\oint\frac{dw}{2\pi i}\frac{f(w)}{\prod_{l}(w-\lambda_l)}
\end{equation}
This is applied to the matrices (\ref{latticefred}), by considering the variables $\lambda_a, \lambda_b$ as two complex variables $w,v$ integrated on a closed contour. In order to do so, one employs the function:
\begin{equation}  \label{Adef}
A(w) =  \prod_{l=1}^M \frac{\sinh\frac{\gamma}{\pi}(w-\mu_l)}{\sinh\frac{\gamma}{\pi}(w-\mu_l-i\pi)}
		\frac{\sinh\frac{\gamma}{\pi}(w-\lambda_l-i\pi)}{\sinh\frac{\gamma}{\pi}(w-\lambda_l)}
\end{equation}
The latter expression contains both the poles in the values of the $\lambda$ roots to be summed over and the zeros in the values of the $\mu$ roots to be avoided, a fact that allows us to keep the contour of integration under control for every couple of states. Be the reader aware that we are using rescaled variables. 

If we consider first the state $\{\mu\}$ to be associated with a (twisted) excited state and the state $\{\lambda\}$ with the ground state, it is sufficient to consider a contour that encircles the real axis. We underline that all the zeros of the factor
\begin{equation}\label{BtoZ}
1+B^{\omega}_\mu(\lambda_a) \to 1+e^{iZ_\mu^{\omega}(w)}
\end{equation}
are all the real roots and holes in the state $\{\mu\}$. This means that the product 
$$\prod_l\sinh\frac{\gamma}{\pi}(w-\mu_l)$$
will cancel all the poles corresponding to real roots, but not the ones corresponding to holes, which will be treated separately.

With $w$ a generic complex variable, having $0<|\Im\,w|<\pi\min(1,p)$ strictly, it is possible to exponentiate the product and apply the formula (\ref{sums}) to the sum of logarithms. It is however necessary to choose the contour in a way to avoid the branch cuts, which is simply done, also numerically, provided $\gamma$ is not too close to $\pi$ or to zero. The result is:
\begin{eqnarray}\label{A}
\qquad\quad A(w)&=&
\frac
{\prod_{c}\sinh\frac{\gamma}{\pi}(w-\lambda_c-i\pi)\prod_{h}\sinh\frac{\gamma}{\pi}(w-\lambda_h)}
{\prod_{c}\sinh\frac{\gamma}{\pi}(w-\lambda_c)\prod_{h}\sinh\frac{\gamma}{\pi}(w-\lambda_h-i\pi)}
\nonumber\\ 
&&
\exp\Big[i\frac{\gamma}{\pi}\sum_{\sigma=\pm}\sigma\int_{-\infty}^{\infty}\frac{dx}{2\pi}
\left(\coth\frac{\gamma}{\pi}(w-x^\sigma-i\pi)-\coth\frac{\gamma}{\pi}(w-x^\sigma)\right)
\mathcal{L}^\sigma(x) \nonumber\\
&&
+\int_{-\infty}^{\infty} dx \sum_\alpha c_\alpha G(\lambda_\alpha-x) \log \frac{\sinh\frac{\gamma}{\pi}(x-w-i\pi)}{\sinh\frac{\gamma}{\pi}(x-w)}\Big]
\end{eqnarray}
which holds whenever $w$ is not on the real axis. However, it is most simply written if one shifts the argument by $i\pi/2$, which yields (\ref{Aplus}) and can be used to compute the factors needed in (\ref{kernelmulambda}), provided $ p \ne 1/2 $. The form above, instead, seems to be more useful for numerics and for extracting the expression $ResA(\lambda_c)$ in (\ref{kernellambdamu}).

If the counting function in (\ref{BtoZ}) refers to an excited state, then it is necessary to subtract from the sum over poles the unwanted ones corresponding to holes, a task which is performed by using a term like:
\begin{equation}\label{holesremover}
\frac{\gamma}{\pi}\sum_{holes}\frac{1}{\sinh\frac{\gamma}{\pi}(w-\lambda_h)}\frac{A(\lambda_h)}{2\pi\,Z_{\lambda}'(\lambda_h)}K(\lambda_h-v)
\end{equation}
as long as the number of holes is of order of unity.

It is also possible to extend our analysis to the case of $\{\mu\}$ being the ground state and $\{\lambda\}$ an excited state: holes need not to be subtracted anymore (the ground state has all the Dirac sea filled), but complex roots outside the contour need to be explicitly added when computing the Fredholm determinant.

This cannot always be done by deforming contours, because of the poles in the kernel at $u-v = \pm\pi,\pm \pi p$. It then follows that roots lying beyond $\min(\frac{\pi}{2},\frac{\pi p}{2})$ must be treated separately and enclosed in different contours.

Taking into consideration neutral states with rapidities quantized with half-integers \cite{FRT98b,FRT98c} the close pairs of complex roots approach their infinite-volume position (\ref{array-I-even}) keeping their distance larger than $\pi$. Moreover, as was observed in Section \ref{sec:SGvsXXZ}, the wide roots do not correspond to any creation operator in the attractive regime, since their presence does not modify the total spin. Then, in the attractive regime, the prescription for the external contours is to surround the region whose imaginary part is $\frac{\pi}{2}<|\Im\theta|<\pi p$. On the other hand, it is known that the antisymmetric soliton-antisoliton states are described by a pair of close roots, so once again, the prescription applies.

The Fredholm determinants are computed on a contour:
\begin{equation}\label{contourfred}
\det\left[\mathbf{1}-\frac{\gamma}{\pi}\hat O \right]=\exp\left\{
\sum_n\frac{1}{n}\left(\frac{\gamma}{\pi}\right)^n\oint\frac{dv_1}{2\pi i}\ldots\oint\frac{dv_n}{2\pi i}
\hat O(v_1,v_2)\hat O(v_2,v_3)\ldots \hat O(v_n,v_1)\right\}
\end{equation}
The contours have to surround all the roots on the real axis; moreover, for excited states, they also need to encircle the complex roots. In principle, one can surround each root by a small circle, taking care to avoid that two points in contours are separated by $\frac{\pi p}{2},i\frac{\pi p}{2}$. This is the prescription for the repulsive regime.

In the attractive regime, the contours encircle only close roots, with $|\Im\theta|<\pi\,p$. In particular, it is known that the pair of close roots describing the polarization of a soliton-antisoliton pair, has an imaginary part which reaches the values $\pm i\frac{\pi}{2}$ from above. The same arguments extend to all the close roots quantized with half-integers. Hence, our contours surround the region of the complex plane $\frac{\pi}{2}<|\Im\theta|<\pi p$.

\begin{figure*}
  \begin{center}
\includegraphics[width=0.45\textwidth]{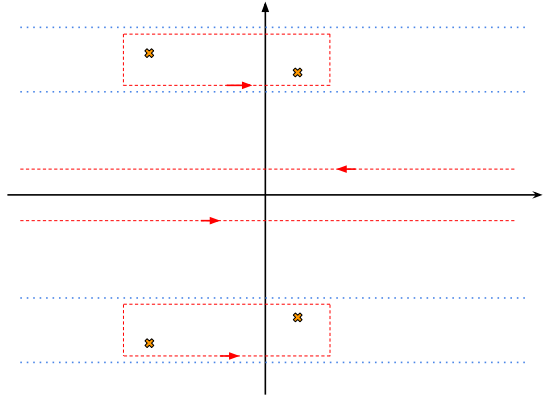}
\caption{contours surrounding roots of the Fermi sea on the real axis and complex roots for $p<1$. The dotted lines are at $\Im\theta=\pm\frac{\pi}{2},\pm\pi p$.}
 \end{center}
\end{figure*}

Indicating by a subscript the states that enter in the matrix element, the overlap kernels, as function of the field theory (rescaled) rapidities, are:
\begin{eqnarray}\label{Ucont}
U_{\lambda,\mu}(x,y)&=&\frac{1}{2\pi i}\frac{A(x)}{1+e^{i Z_0(x)}} K_{-\omega}(x-y)	\nonumber\\
U_{\mu,\lambda}(x,y)&=&\frac{1}{2\pi i}\frac{A(x)^{-1}}{1+e^{i Z_\lambda(x)}} K_{\omega}(x-y)
-\frac{\gamma}{\pi}\sum_{h}\frac{1}{\sinh(\frac{\gamma}{\pi}(x-\lambda_h))}
\frac{A(\lambda_h)^{-1}}{2\pi Z'_\lambda(\lambda_h)}
K_{\omega}(\lambda_h-y)
\nonumber\\
\end{eqnarray}
and the norm kernels
\begin{eqnarray}\label{Unormcont}
U_\mu(x,y) &=& \frac{1}{2\pi i}\frac{1}{1+e^{iZ_0(x)}}K(x-y)   \nonumber\\
U_\lambda(x,y) &=& \frac{1}{2\pi i}\frac{1}{1+e^{iZ_\lambda(x)}}K(x-y)
-\frac{\gamma}{\pi}\sum_{h}\frac{1}{\sinh(\frac{\gamma}{\pi}(x-\lambda_h))}\frac{1}{2\pi Z_\lambda'(\lambda_h)}K(\lambda_h-y)
\nonumber\\
\end{eqnarray}
The previous expressions hold for $w$ not on the real axis. They are valid also in the scaling limit but, if the limit is taken at this stage, the integrations with the kernels above are diverging. Therefore, we need to manipulate further these expressions, which will be done in the following section.

\subsection{More on determinants} \label{sec:moredet}
The integral operators in (\ref{Ucont},\ref{Unormcont}) above are not of trace class when the integration is performed over the whole real axis. To proceed with the numerical evaluation, one would be forced to introduce a cutoff; the function $A$ above, which is present in the overlap integrals, tends exponentially to unit value when its argument has large real part, for any couple of the states. It follows that the asymptotic behaviour for large rapidities of the ``overlap`` kernels is the same. Then, by considering normalized overlaps as in (\ref{amplitudedef}), one may argue that the result is independent from the cutoff and the ratio of determinant such as (\ref{contourfred}) is finite.

However, we will not need to follow this path. Instead, we start from integrations which surround the real roots of an eigenstate of the finite chain, so that contour integrations must surround only a finite number of roots. Then, we split integrations and group them into series. Finally, the boundaries of integration are sent to infinity, providing an exact summation of these families of terms when the scaling limit is taken. As a result, we retain a Fredholm determinant form, but with a trace class kernel.

Considering the kernel in the expression of the norm, the integral on the contour can be written as:
\begin{equation}\label{contourtoline}
\oint\frac{dw}{2\pi i}\frac{K(w-v)}{1+e^{i Z_{\lambda}(w)}}=
\int_{-\Lambda}^\Lambda \frac{dw}{2\pi i}\left(\frac{K(w^+-v)}{1+e^{-iZ_{\lambda}(w^+)}}+\frac{K(w^--v)}{1+e^{iZ_{\lambda}(w^-)}}-K(w^+-v))\right)
\end{equation}
where the integration boundaries are intended to be more negative and more positive than the real roots with largest absolute values, and the parts of the contour which are perpendicular to the real axis are assumed to give negligible contribution.
 By writing down explicitly the first terms of the Fredholm determinant expansion with this form for the kernel and noting that the function $K$ has no poles on the real axis, it is possible to show (see appendix \ref{app:kernel}) that:
\begin{eqnarray}\label{ktog}
\sum_n\frac{1}{n}\oint\frac{dw_1}{2\pi i}\ldots\oint\frac{dw_n}{2\pi i}\frac{K(w_1-w_2)}{1+e^{i Z(w_1)}}\ldots\frac{K(w_{n}-w_1)}{1+e^{i Z(w_n)}}
\quad \to \quad
\tr\log\left[1-\frac{i\hat K}{2\pi}\right] \nonumber\\
\qquad
+\sum_n\frac{1}{n}\sum_{\sigma_1\ldots\sigma_n=\pm}\intop_{-\infty}^\infty \frac{dw_1}{2\pi }\ldots\intop_{-\infty}^\infty \frac{dw_n}{2\pi } 
\left( \frac{G(w_1^{\sigma_1}-w_2^{\sigma_2})}{1+e^{-iZ(w^{\sigma_1})}} \ldots \frac{G(w_n^{\sigma_n}-w_2^{\sigma_1})}{1+e^{-iZ(w^{\sigma_1})}}\right)
\end{eqnarray}
for some generic counting function $Z$, where we have used the fact that the kernel in the NLIE (\ref{NLIE}) can be expressed \cite{DDV87} in terms of the spin chain kernel as:
\begin{equation}\label{gseries}
G(w)=\sum_{n=1}^\infty (-1)^{n-1} (-i K)^{\star n}(w) = \intop_{-\infty}^\infty dq e^{iwq}\frac{-i\hat K(q)}{1-i\hat K(q)}
\end{equation}
in which the superscript $\star n$ denotes $n$-times convolution. 

The first term in the right-hand side of (\ref{ktog}) is formally divergent. However, it can be combined with the analogous terms arising from the overlap determinant and written as a series, as seen in Section \ref{sec:normalization}.

In order to deal with the overlap determinant, one can similarly set:
\begin{equation}\label{gomega}
G_\omega(w)=\sum_{n=1}^\infty (-1)^{n-1} (-i K_\omega)^{\star n}(w) = \intop_{-\infty}^\infty dq e^{iwq}\frac{-i\hat K_\omega(q)}{1-i\hat K_\omega(q)}\;
\end{equation}
 The Fourier transform is performed by separating the asymptotic behaviour of the function and by regularizing through a damping exponential.
A reasoning analogous to that for the norm leads to the following overlap determinant:
\begin{eqnarray}\label{nomrdetone}
&&\det\left[1-\sum_{\sigma,\sigma'=\pm}  W_{\sigma\sigma'}^\lambda(w,v)\right] 
\end{eqnarray}
where we have defined a function $W$ of two real arguments and with two contour subscripts. To be clear, we denote, e.g., $W^\lambda_{+-}(x,y)$ the expression resulting from the dressing of the kernel of the overlap determinant, where the first variable has real part $x$ and is on the upper branch of the contour and the second has real part $y$ and is on the lower one. Explicitly:
\begin{eqnarray}\label{nomrdetoneker}
 W_{\sigma\sigma'}^\lambda(w,v)&=&\frac{1}{2\pi} \left(\frac{A(w^\sigma)}{1+e^{-\sigma i Z_0(w^\sigma)}}
-\delta_{\sigma,+}\left(A\left(w^+\right)-1\right)\right)
G_{-\omega}(w^\sigma-v)
\end{eqnarray}
 which is manifestly of trace class due to the presence of the soliton and antisoliton filling fractions and to the asymptotic behaviour of the function $A$, which tends to one for large values of the argument. 

We will shift the contours of integration to a definite value $\Im w = \eta$ and define the kernel $W$ with real argument only.
\begin{eqnarray}
W_{\lambda,0}^{\sigma,\sigma'}(w,v)&=&
\frac{1}{2\pi}\left(\frac{A(w+i\sigma\eta)}{1+e^{-i\sigma Z_0(w+i\sigma\eta)}}-\delta_{\sigma+}\left(A_+(w)-1\right)\right) 
G_{-\omega}\left(w-v+i({\sigma-\sigma'})\eta \right)  \nonumber\\
W_{\lambda,0}^{\sigma'}(\lambda_c,v)&=&
 \frac{Res A(\lambda_c)}{1+e^{iZ_0(\lambda_c)}}
G_{-\omega}\left(\lambda_c-v-i\sigma'\eta \right)  \nonumber\\
W_{\lambda,0}^{\sigma}(w,\lambda_c)&=&
\frac{1}{2\pi}\left(\frac{A(w+i\sigma\eta)}{1+e^{-i\sigma Z_0(w+i\sigma\eta)}}-\delta_{\sigma+}\left(A_+(w)-1\right)\right)
G_{-\omega}\left(w-\lambda_c+i\sigma\eta\right)  \nonumber\\
W_{\lambda,0}(\lambda_c,\lambda_{c'})&=&
\left( \frac{Res A(\lambda_c)}{1+e^{iZ_0(\lambda_c)}}-1\right)
G_{-\omega}\left(\lambda_c-\lambda_{c'}\right)
\end{eqnarray}
\begin{eqnarray}
W_{0,\lambda}^{\sigma,\sigma'}(w,v)&=&
\frac{1}{2\pi}\left(\frac{A^{-1}(w+i\sigma\eta)}{1+e^{-i\sigma Z_{\lambda}(w+i{\sigma}\eta)}}-\delta_{\sigma+}\left(A^{-1}_+(w)-1\right)\right)
G_{\omega}\left(w-v+i(\sigma-\sigma')\eta \right)  \nonumber\\
W_{0,\lambda}^{\sigma'}(\lambda_h,v)&=&
-\frac{A^{-1}(\lambda_h)}{Z_\lambda'(\lambda_h)}G_{\omega}\left(\lambda_h-v-i\sigma'\eta\right)  \nonumber\\
W_{0,\lambda}^{\sigma}(w,\lambda_h)&=&
\frac{1}{2\pi}\left(\frac{A^{-1}(w+i\sigma\eta)}{1+e^{-i\sigma Z_{\lambda}(w+i{\sigma}\eta)}}-\delta_{\sigma+}\left(A^{-1}_+(w)-1\right)\right)
G_{\omega}\left(w-v+i(\sigma-\sigma')\eta\right)  \nonumber\\
W_{0,\lambda}(\lambda_h,\lambda_{h'})&=&-
\left(\frac{A^{-1}(\lambda_h)}{Z_\lambda'(\lambda_h)}-1\right)G_{\omega}\left(\lambda_h-\lambda_{h'}\right)
\end{eqnarray}
It is then possible to apply again the partial summation of some families of terms in the Fredholm determinant series along the lines of appendix \ref{app:kernel}. In this case, the terms that can be summed over are the ones with domain only in the upper branch, i.e., those like
\begin{equation}
\frac{1}{2\pi}\delta_{\sigma+}\left(A^{-1}_+(w)-1\right)G_{\omega}\left(w^{\sigma}-v^{\sigma'}\right)
\end{equation}
 Unfortunately, it is not as easy as in the previous case to Fourier transform and sum over all the terms, hence the result is given in the rather involved form (\ref{Fomega}).

 The expressions above imply explicit subtraction of holes in the ket and summation of the complex roots of the bra, when used to compute the determinants of (\ref{Rterm}). For instance, when multiplying $W_{\lambda,0}$ with itself, one has:
\begin{equation}\label{Wsquare}
\left(W_{\lambda,0}^2\right)^{+,-}(w,v) = 
\sum_{\sigma=\pm} \int_{-\infty}^{\infty}dz W_{\lambda,0}^{+,\sigma}(w,z) W_{\lambda,0}^{\sigma,-}(z,v)
+\sum_{\lambda_c} W_{\lambda,0}^{+}(w,\lambda_c) W_{\lambda,0}^{-}(\lambda_c,v)
\end{equation}
and similarly for $W_{0,\lambda}$.
 Then one can make use of
\begin{equation}\label{Freddie}
\det\left[1-\hat W\right] = -\sum_{n=1}^{\infty}\frac{1}{n} \tr\left[\hat W^{n} \right]
\end{equation}
with the above recipe for the product and the trace.

At this stage, contour indexes can be interpreted as species indexes, as they enter in a symmetric way in the result. In order to do so, one could deform the contour up to $|\Im w=\pi/2|$ and, following \cite{DDV95}, define the (complex) soliton and antisoliton pseudoenergies as:
\begin{eqnarray}\label{solitonpseudoenergydef}
 \varepsilon_+(\theta) &=&  -i Z(\theta+i\frac{\pi}{2})  \nonumber\\
 \varepsilon_-(\theta) &=& i Z(\theta-i\frac{\pi}{2})  
\end{eqnarray}
Then we understand the factor containing the exponentiated counting function as a filling fraction. This interpretation is very suggestive of a possible extension to two--point functions of the validity of some sort of Leclair-Mussardo formula \cite{LM99}, provided suitable excited state pseudoenergies are used. To avoid heavier notation, we write the final result in terms of the counting function alone.

\subsection{The prefactor} \label{sec:prefactor}
As a warm-up, we look at the argument of the cosine in (\ref{scal-prod2}). By applying the formula (\ref{sums}), we see that the term proportional to $N$ cancel in the difference and we obtain
\begin{eqnarray}\label{Sigma}
\sum_{j}\left(\lambda_j-\mu_j\right) &=& -\sum_\alpha c_\alpha \left(\lambda_\alpha- \int_{-\infty}^{\infty} du u G(\lambda_\alpha-u) \right) \nonumber\\
&&-\sum_{\sigma=\pm}\int_{-\infty}^{\infty} \frac{du}{2\pi i} \left(u^\sigma -\int dx G(u^\sigma-x) x \right)
 \log'\frac{1+e^{i\sigma Z_{\lambda}(u^\sigma)}}{1+e^{i\sigma Z_{0}(u^\sigma)}} 
\end{eqnarray}
with $\lambda_\alpha$ being, as usual, the roots that define the source terms in (\ref{NLIE}). Then, we integrate by parts the second term and arrive to the expression (\ref{coshfactor}), using the asymptotic behaviour $\chi(\pm\infty)=\pm\frac{\pi}{2}\frac{p-1}{p}$ \cite{DDV93}.

We now turn to the analysis of the factor (\ref{Sfactordef}). Here the complication lies in the double product, but there are no conceptual difficulties in exponentiating this expression and performing the sum of the resulting logarithms with the procedure described above, since the arguments of the logarithms never cross the cut. The first step is
\begin{eqnarray}
&& \sum_{j,k}\left(\log\frac{\sinh(\lambda_j-\mu_k-i\gamma)}{\sinh(\lambda_j-\lambda_k-i\gamma)}+
\log\frac{\sinh(\mu_j-\lambda_k-i\gamma)}{\sinh(\mu_j-\mu_k-i\gamma)}\right) \nonumber\\
&& = \sum_j\Big\{ -\sum_{\sigma=\pm}\frac{\sigma}{2\pi i}\int_{-\infty}^\infty dx \left(\Delta\star\log'\frac{1+e^{i\sigma Z_0}}{1+e^{i\sigma Z_{\lambda}}}\right)
     \log\frac{\sinh(\lambda_j-x^\sigma-i\gamma)}{\sinh(\mu_j-x^\sigma-i\gamma)}\nonumber\\
&& + \int_{-\infty}^\infty du \hat\Delta(u)\log\frac{\sinh(\lambda_j-u-i\gamma)}{\sinh(\mu_j-u-i\gamma)}
\end{eqnarray}
where both counting functions are relative to a finite number of rapidities. Then one is to apply again (\ref{sums}) to the $j$ index. After passing to the scaling limit (\ref{scalingDdV}) we find the result (\ref{Csfactor}), with the counting functions satisfying (\ref{NLIE}).

The complex roots are, by definition, away from the real axis. Since in the product defining the complex root factor (\ref{complfacprod}) there appear all and only the real solutions, it is natural to consider the logarithm and chose a suitable contour around the real axis to perform the sum over roots. The resulting expression is given in (\ref{c-factor}).

We turn to the analysis of the factor containing the position of the holes. The counting function itself may be non monotonic in some region of the real axis for some class of states and at small volumes $mL\sim O(1)$, a circumstance which is connected with the appearance of \emph{special} roots when the counting function also crosses a quantization point within that region. We assume that this is not the case, even if, in principle, it can be worked around by splitting the sum over different regions in which the counting function is monotonically increasing. Having written the product (\ref{holefacprod}) in terms of positive functions, we can now take the logarithm and perform the scaling limit. 

Such limit is somewhat simplified by the fact that the sum is performed only over real solutions of the (\ref{polesfactor}). It can be performed by standard techniques, but some details are in order. After applying the summation procedure, one of the terms has the form
\begin{eqnarray}
\mathcal{I}_0=\int_{-\infty}^\infty \frac{du}{2\pi}\left(\frac{N}{\cosh(u-\Theta)}+\frac{N}{\cosh(u+\Theta)}\right)\log\frac{\varphi_0(x,u)}{\varphi_{\lambda}(x,u)}
\end{eqnarray}
This integral is similar to the ones needed for the computation of energy and momentum (\cite{DDV97}, see also \cite{Fth}): one uses the fact that the counting function of a finite chain has a well defined limit when the argument is sent to infinity
\begin{equation}\label{Zasymptotics}
\left\{
\begin{array}{ccc}
Z(+\infty) & = & N\pi +\frac{p-1}{p+1}\pi S +2\pi \sign(p-1) M_{W\downarrow} +2\omega \\
Z(-\infty) &  =& -N\pi-\frac{p-1}{p+1}\pi S -2\pi \sign(p-1) M_{W\uparrow} +2\omega
\end{array}
\right.
\end{equation}
where $M_{W\downarrow}$ and $M_{W\uparrow}$ are for the number of wide roots below and above the real axis. It follows that:
\begin{equation}\label{I0}
\mathcal{I}_0 = \sign(p-1) M_{W}
\end{equation}
This integral enters both in the evaluation of the ``hole'' factor and in the product (\ref{Phifactordef}). Here below, we report the result of the summation on the first index:
\begin{eqnarray}
&&\sum_{j,k}\log
\frac{\varphi_{\lambda}(\tilde\lambda_l,\tilde\lambda_k)\varphi_0(\mu_j,\mu_k)}
{\varphi_{\lambda}(\tilde\lambda_l,\mu_k)\varphi_0(\tilde\lambda_j,\mu_k)}=  
\nonumber\\
&& \qquad
\sum_j \Big\{
\sum_{\sigma=\pm} 
\int_{-\infty}^{\infty}
\frac{du}{2\pi i \sigma}\log'\frac{1+e^{i\sigma Z_{\lambda}(u^\sigma)}}{1+e^{i\sigma Z_0(u^\sigma)}} 
\int_{-\infty}^{\infty} dx \Delta(u-x) \log\frac{\varphi_0(\tilde\mu_j,x^\sigma )}{\varphi_{\lambda}(\tilde\lambda_j,x^\sigma )}  \nonumber\\
&& \qquad
+\sum_\alpha c_\alpha \int_{-\infty}^{\infty} du G(\lambda_\alpha-u)\left( \log\varphi_{\lambda}(\tilde\lambda_j,u^-) - \log\varphi_0(\tilde\mu,u^-) \right)
\Big\}
\end{eqnarray}
 the second step, again by taking into account the asymptotic behaviour (\ref{Zasymptotics}), can be analogously performed and yields (\ref{Phifactor}) as a result.

\subsection{A note on normalization}\label{sec:normalization}
The ``normalized'' amplitudes $\tilde\mathcal{A}$ are defined from the bare amplitudes (\ref{amplitudedef}) by dividing by the factor
\begin{equation}\label{a_omega}
a_\omega = \frac{\det\left[1+K_\omega\right]\det\left[1+K_{-\omega}\right]}{\det\left[1+K\right]^2}
\end{equation}
with kernel (\ref{Komegadef}) 
and $x$ on the real axis. The $K_{\omega}$ operators are not of trace class and we have not found a way of directly evaluating these determinants up to present. Moreover, at first sight, it appears that the factor (\ref{a_omega}) can be written in terms of a difference of Fredholm series, whose terms are one by one divergent.

However, a definite expression for this factor, in the form of a series, can be given considering (\ref{Gseries}) for $x=0$, for which the operator (\ref{expfield}) reduces to the expectation value of the identity on the finite--size vacuum. Therefore, imposing
\begin{equation}\label{identityvev}
\left\langle \mathbf{I} \right\rangle_L = 1
\end{equation}
for every size $L$, we obtain the exact normalization in the form (\ref{Nseries}). 

Let us add that, being the series for the generating function derived from (\ref{latticeamplitudexpansion}), we expect similar convergence properties for (\ref{Gseries}) directly. However, it seems that this issue should be tackled by extensive numerical analysis.

\section{Conclusions}\label{sec:conclusions}
We have presented an exact expression for the generating function of connected correlation functions on a cylinder, where the compactified direction is space, for the sine-Gordon quantum field theory.

To take into account corrections which enter as exponentials in the size, the knowledge of the spectrum and of the form factors of the theory in infinite volume is not enough. To circumvent this problem, the computations were carried on in the framework of the Destri-De Vega lattice regularization, formulated in terms of an inhomogeneous XXZ spin chain.

In this framework, the problem was similar to the study of the expectation value of the magnetization of the spin chain in a given interval: due to the available results for this model, the most important of which is the solution of the quantum inverse scattering, we managed to write the exact vacuum expectation value of the generating function in the form of a series in (\ref{Gseries}) and performed the appropriate scaling limit of each term to obtain the result (\ref{amplitude}).

Each term in our expansion is associated to one of the states of the field theory in finite volume and can be interpreted as a generating function of form factors. The determination of the exact states relies on the ability of solving self-consistently the Destri-De Vega nonlinear integral equation for all the allowed source terms, which is in general a difficult task and limits for the moment the practical applicability of the method to some classes of finite volume states.

 Having shown its relevance in the computation of correlation functions, we hope to be able to extend the analysis of the nonlinear integral equation in the future. Further work would be moreover required to explore the advantages and limitations of this kind of formalism in the actual computation of correlation functions, as well as to identify the features which may be generally valid for other integrable field theories.

\section*{Acknowledgments}
I am grateful to G. T\'ak\'acs for his help and support, including a review of this manuscript. I also thank \mbox{F. Ravanini} for valuable lessons and recommendations, as well as C. Matsui and B. Pozsgay for fruitful and pleasant discussions and A. Trombettoni for his suggestions during the preparation of this revised version.

\appendix

\section{Manipulation of the matrix elements} \label{app:contoursum}
To compute the determinant, we multiply the matrix $H$ by a conveniently defined matrix $M$ and its inverse. In the case of a twisted state defined by the roots $\{\lambda\}$ and the state defined by the roots $\{\mu\}$, we consider the matrix
\begin{equation}
M_{j,k}=\frac{\cosh(\mu_j-\lambda_k) \prod_{l\ne k} \sinh(\mu_j-\lambda_l)}{\prod_{l\ne j} \sinh(\mu_j-\mu_l)}
\end{equation}
whose elements are $i\pi$-antiperiodic functions of the rapidity $\mu_j$. Its determinant is 
\begin{equation}
\det M =\prod_{j<k}\frac{\sinh(\lambda_j-\lambda_k)}{\sinh(\mu_j-\mu_k)}\cosh(\sum_{l}\lambda_l-\sum_{l}\mu_l)
\end{equation}
Then the matrix product 

$\tilde H_{j n}M_{n k}$
can be computed by considering the integral
\begin{eqnarray}
\oint\frac{dw}{2\pi i} \frac{\sinh(-i\gamma)}{\sinh(w-\lambda_j)\sinh(w-\lambda_j\pm i\gamma)}
            \frac{\cosh(w-\lambda_k) \prod_{l\ne k} \sinh(w-\lambda_l)}{\prod_{l} \sinh(w-\mu_l)}
\end{eqnarray}
which is vanishing when the contour of integration surrounds the real axis and the strip $[-\pi/2,\pi/2]$ along the imaginary axis.

Then the result of the matrix multiplication is
\begin{eqnarray}\label{latticefred}
\left[H^{\omega}\cdot M\right]_{a,b}=	
(-e^{-2i\omega}) d(\lambda_a) \left(1+B^{\omega}_\lambda(\mu_a)\right) 
\prod_l\sinh(\lambda_l-\mu_a+i\gamma)
\frac{\prod_{l\ne a}\sinh(\mu_a-\mu_l)}{\prod_{l}\sinh(\mu_a-\lambda_l)}
 \nonumber\\
\Big\{
\delta_{ab}-\frac{\prod_{l}\sinh(\mu_a-\lambda_l)}{1+B^\omega_\lambda(\mu_a)}
\frac{1}{\prod_{l\ne a}\sinh(\mu_a-\mu_l)}
\frac{\sinh(\mu_a-\mu_l-i\gamma)}{\sinh(\mu_a-\lambda_l-i\gamma)}
\nonumber\\
\big(\coth(\mu_a-\mu_b-i\gamma) -e^{2i\omega}\coth(\mu_a-\mu_b+i\gamma) \big)
\Big\}
\end{eqnarray}
from which the Fredholm determinant in the limit in which the size of the matrices goes to infinity can be recovered.

\section{Dressing of the kernels}\label{app:kernel}
We start from a situation in which the number of roots is finite. Given an integral operator $Q$ and the kernel $K$, we have that:
\begin{equation}
\det\left[1-\left(Q-K\right)\right] = \det\left[1-Q*\frac{1}{1+K}\right] \det\left[{1+K}\right] 
\end{equation}
 Here, the symbol $*$ is used to denote a convolution in which the extrema of integration are not to infinity, yet they are large enough to contain all the real roots.
 The equality can be checked by taking the logarithm of the above expression, expanding and reordering terms:
\begin{eqnarray}
 \tr\Big[\log\left[1-\left(Q-K\right)\right]\Big]
=\tr\Big[K - \frac{1}{2} K^{*2} + \frac{1}{3} K^{*3} - \frac{1}{4} K^{*4} + \frac{1}{5}K^{*5} -\ldots 
 \nonumber\\
-\left(Q-Q*K+Q*K^{*2}-Q*K^{*3}+Q*K^{*4} \ldots \right)\nonumber\\
-\frac{1}{2}\left(Q^{*2}-2Q^{*2}*K+2Q^{*2}*K^{*2}+(Q*K)^{*2}-2 (Q*K)^{*2}*K
 \ldots \right)\nonumber\\
 -\frac{1}{3}\left(Q^{*3}-3Q^{*3}*K+3Q^{*3}*K^{*2}+3Q*K*Q^{*2}*K+ \ldots \right)\nonumber\\
 -\frac{1}{4}\left(Q^{*4}-4Q^{*4}*K + \ldots \right) \Big]\nonumber\\
 - \ldots \nonumber\\
=\tr\left[ \log\left(1-Q*\left(1-K+K^{*2}-K^{*3}+\ldots \right)\right)\right]
 +\tr\left[\log\left(1+K\right)\right]
\end{eqnarray}
which is used both for the norm and for the overlap kernel, with different $Q$s and either $K_\omega$ or $K_0$. Obviously, the symbol $K^{*3}$ means $K*K*K$ and so on.

After organizing the summation into series, it is safe to take the scaling limit in the first term of the last expression, since it is of trace class on the real axis. Then the $*$ becomes a convolution over the whole real axis ($\star$). For what the second term is concerned, this becomes formally divergent as the integration boundaries go to infinity. Nevertheless, these terms are independent of the state and can be factorized out of the sum (\ref{Gseries}), grouped together and represented as a series, as explained in Section \ref{sec:normalization}. In other words, there is no need to evaluate them.

To perform the Fourier transform of $K_\omega$, it is simpler to use the Bethe rapidities and write the kernel as:
\begin{equation}\label{bKomega}
K_\omega(\lambda)=e^{i\omega}\left[ \cos\omega K_0(\lambda) -i\sin\omega \left(K_+(\lambda)+2\tanh(\lambda)\right) \right]
\end{equation}
where
\begin{eqnarray}\label{Kplus}
K_+(\lambda)=\coth(\lambda-i\gamma)+\coth(\lambda+i\gamma)-2\tanh(\lambda)
\end{eqnarray}
By applying the residue theorem, one can see that
\begin{equation}\label{Ft}
\hat K_0(k) = i \frac{\sinh\left(\frac{\pi}{2}-\gamma\right)k}{\sinh\frac{\pi\, k}{2}}
\;,\qquad
\hat K_+(k) = -i\frac{\cosh\left(\frac{\pi}{2}-\gamma\right)k -1}{\sinh\frac{\pi\, k}{2}}
\end{equation}
Moreover, we write:
\begin{equation}
\int_{-\infty}^{\infty} d\lambda e^{-i\lambda k} 2\tanh\lambda = - 2 \lim_{\alpha \to 0}\int_{0}^{\infty}  d\lambda (e^{i\lambda k}- e^{-i\lambda k})\tanh\lambda e^{-\alpha k}
\end{equation}
then we make use of the integral representation
\begin{equation}\label{digammaint}
\psi(z)=\int_0^{\infty} dt \left(\frac{e^{-t}}{t} -\frac{(t+1)^{-z}}{t}\right)
\end{equation}
valid for $\Re(z)>0$, to write the integral above as a sum of digamma functions. After further massaging and the limit $\alpha\to 0$, 
we arrive at the expression
\begin{equation}\label{-iK}
\hat K_{\omega}(k) = i e^{i\omega}\frac{\cosh\left( \left(\frac{\pi}{2}-\gamma \right)k + i \omega \right)}{\sinh\frac{\pi\,k}{2}}
\end{equation}
from which (\ref{Gomega}) follows.

\section{Translations}\label{translations}
Within the formalism of the light-cone lattice (see \cite{DDV87,F96lh}), the operators generating translations along the space and time directions can be constructed by successive application of the inhomogeneous transfer matrix at special values of the argument:
\begin{equation}
 e^{-i\frac{a}{2}(E-P)}\left|\Psi(\{\lambda\})\right\rangle = U_L \left|\Psi(\{\lambda\})\right\rangle 
= \hat \tau(\Theta)\left|\Psi(\{\lambda\})\right\rangle = \tau(\Theta|\{\lambda\})\left|\Psi(\{\lambda\})\right\rangle
\end{equation}
\begin{equation}
 e^{-i\frac{a}{2}(E+P)}\left|\Psi(\{\lambda\})\right\rangle = U_R \left|\Psi(\{\lambda\})\right\rangle 
= {\hat \tau(-\Theta)}^\dagger\left|\Psi(\{\lambda\})\right\rangle = \tau(-\Theta|\{\lambda\})^*\left|\Psi(\{\lambda\})\right\rangle
\end{equation}
Considering two neighboring sites with opposite values of the real part of the inhomogeneity:
\begin{eqnarray}
 \frac{e^{-i\omega}\tau_\omega(\Theta|\{\lambda\})}{\tau_0(\Theta|\{\mu\})}
\frac{e^{-i\omega}\tau_\omega(-\Theta|\{\lambda\})}{\tau_0(\Theta|\{\mu\})} 
=e^{-i\, a (\mathcal{P}(\{\lambda\}_\omega)-\mathcal{P}(\{\mu\}))}
\end{eqnarray}
By looking then at the full product in (\ref{latticeamplitudexpansion}), one recovers the expression of momenta analyzed in \cite{Fth}, in presence of a twist and reported in (\ref{DressedMomentum}). An analogous path can be followed by applying the transfer matrices
\begin{equation}
\hat \tau(\Theta)\hat \tau(-\Theta)^\dagger
\end{equation}
and performing the scaling limit on their eigenvalues, in order to obtain translations along the time direction.

\bibliographystyle{nar} 
\bibliography{TESI.bib} 
\end{document}